%Paper: astro-ph/9401039
%From: Tanmay Vachaspati <TVACHASP@PEARL.TUFTS.EDU>
%Date: Sun, 23 Jan 1994 11:58:55 -0400 (EDT)

%%%%%%%%%%%%%%%%%%%%%%%%%%%%%%%%%%%%%%%%%%%%%%%%%%%%%%%%%%%%%%%%%%%%%
%%              JNL.TEX                         Doug Eardley
%%
%%      This is a set of TeX 82 macros designed to produce scientific
%%      papers with a minimum of fuss and using as much of plain.tex as
%%      possible.  The user need only know what is in the TeXbook, and
%%      the macros under ``user definitions'' below.  Also, the user
%%      definitions are intended to be as simple as possible, so that
%%      the user may change them as desired.  I have tried to avoid all
%%      cleverness, although it may have snuck in here and there.
%%
%%      For documentation, see the file JNLHLP.TEX.  Optional features are
%%      contained in the files PPT.TEX (for two-up preprints), REFORDER.TEX
%%      (automatic numbering of references), EQNORDER.TEX (automatic numbering
%%      of equations), and TABLEOFC.TEC (automatic generation of table of
%%      contents).

%%
%%  Font definitions suitable for the IMAGEN-480 (Written by Tony Kennedy)
%%      Define a whole menagerie of pseudo-12pt fonts

\font\twelverm=cmr10  scaled 1200   \font\twelvei=cmmi10  scaled 1200
\font\twelvesy=cmsy10 scaled 1200   \font\twelveex=cmex10 scaled 1200
\font\twelvebf=cmbx10 scaled 1200   \font\twelvesl=cmsl10 scaled 1200
\font\twelvett=cmtt10 scaled 1200   \font\twelveit=cmti10 scaled 1200
\font\twelvesc=cmcsc10 scaled 1200
% \font\twelvesf=cmssmc10 scaled 1200
\skewchar\twelvei='177   \skewchar\twelvesy='60

%  Define \...point macros to change fonts and spacings consistently

     %       Melott
\def\twelvepoint{\normalbaselineskip=12.4pt plus 0.1pt minus 0.1pt
  \abovedisplayskip 12.4pt plus 3pt minus 9pt
  \belowdisplayskip 12.4pt plus 3pt minus 9pt
  \abovedisplayshortskip 0pt plus 3pt
  \belowdisplayshortskip 7.2pt plus 3pt minus 4pt
  \smallskipamount=3.6pt plus1.2pt minus1.2pt
  \medskipamount=7.2pt plus2.4pt minus2.4pt
  \bigskipamount=14.4pt plus4.8pt minus4.8pt
  \def\rm{\fam0\twelverm}          \def\it{\fam\itfam\twelveit}%
  \def\sl{\fam\slfam\twelvesl}     \def\bf{\fam\bffam\twelvebf}%
  \def\mit{\fam 1}                 \def\cal{\fam 2}%
  \def\sc{\twelvesc}               \def\tt{\twelvett}
  \def\sf{\twelvesf}
  \textfont0=\twelverm   \scriptfont0=\tenrm   \scriptscriptfont0=\sevenrm
  \textfont1=\twelvei    \scriptfont1=\teni    \scriptscriptfont1=\seveni
  \textfont2=\twelvesy   \scriptfont2=\tensy   \scriptscriptfont2=\sevensy
  \textfont3=\twelveex   \scriptfont3=\twelveex  \scriptscriptfont3=\twelveex
  \textfont\itfam=\twelveit
  \textfont\slfam=\twelvesl
  \textfont\bffam=\twelvebf \scriptfont\bffam=\tenbf
  \scriptscriptfont\bffam=\sevenbf
  \normalbaselines\rm}

%       tenpoint

%%
%%      Various internal macros
%%

\def\beginlinemode{\endmode
  \begingroup\parskip=0pt \obeylines\def\\{\par}\def\endmode{\par\endgroup}}
\def\beginparmode{\endmode
  \begingroup \def\endmode{\par\endgroup}}
\let\endmode=\par
{\obeylines\gdef\
{}}
        %       Melott
        %       Melott
\def\singlespace{\baselineskip=\normalbaselineskip}

\def\oneandahalfspace{\baselineskip=\normalbaselineskip
  \multiply\baselineskip by 3 \divide\baselineskip by 2}
\def\doublespace{\baselineskip=\normalbaselineskip \multiply\baselineskip by 2}

\newcount\firstpageno
\firstpageno=2

\footline={\ifnum\pageno<\firstpageno{\hfil}\else{\hfil\twelverm\folio\hfil}\fi}

\def\toppageno{\global\footline={\hfil}\global\headline
  ={\ifnum\pageno<\firstpageno{\hfil}\else{\hfil\twelverm\folio\hfil}\fi}}

\let\rawfootnote=\footnote              % We must set the footnote style
\def\footnote#1#2{{\rm\singlespace\parindent=0pt\parskip=0pt
  \rawfootnote{#1}{#2\hfill\vrule height 0pt depth 6pt width 0pt}}}
\def\raggedcenter{\leftskip=4em plus 12em \rightskip=\leftskip
  \parindent=0pt \parfillskip=0pt \spaceskip=.3333em \xspaceskip=.5em
  \pretolerance=9999 \tolerance=9999
  \hyphenpenalty=9999 \exhyphenpenalty=9999 }
\def\dateline{\rightline{\ifcase\month\or
  January\or February\or March\or April\or May\or June\or
  July\or August\or September\or October\or November\or December\fi
  \space\number\year}}
\def\received{\vskip 3pt plus 0.2fill
 \centerline{\sl (Received\space\ifcase\month\or
  January\or February\or March\or April\or May\or June\or
  July\or August\or September\or October\or November\or December\fi
  \qquad, \number\year)}}

%%
%%      Page layout, margins, font and spacing        (feel free to change)
%%

\hsize=6.5truein
\hoffset=0.0truein
\vsize=8.5truein
\voffset=0.25truein
\parskip=\medskipamount
%\toppageno
\nopagenumbers
\twelvepoint
\doublespace
\def\\{\cr}
\overfullrule=0pt % delete the nasty little black boxes for overfull box

%%
%%      The user definitions for major parts of a paper (feel free to change)
%%

%\input timestamp\rightline{Draft \timestamp}}  %  "Draft", Timestamp

 % Preprint number at upper right of title page

\def\title#1{                   %  Title on title page
   \null \vskip 3pt plus 0.3fill \beginlinemode
   \doublespace \raggedcenter {\bf #1} \vskip 3pt plus 0.1 fill}

\def\author                     %  Author(s) name(s)  on title page
  {\vskip 3pt plus 0.1fill \beginlinemode \doublespace \raggedcenter}

\def\affil                      % Affiliations (can intermix with \author)
  {\vskip 3pt \beginlinemode \doublespace \raggedcenter \it}

\def\abstract                   % Begin abstract
  {\vskip 3pt plus 0.1fill \subhead {Abstract:}
   \beginparmode \narrower \oneandahalfspace }

\def\endtopmatter               % End title page, begin body of paper
  {\vskip 3pt plus 0.1fill \endpage \body}

\def\body                       % Begin text body;  can be used to end
  {\beginparmode}               % \title, \author, \affil, \abstract,
                                % \reference, or \figurecaption modes

\def\head#1{                    % Head;  NOTE enclose the text in {}
   \goodbreak \vskip 0.4truein  %  e.g., \head{I. Introduction}
  {\immediate\write16{#1} \raggedcenter {\sc #1} \par}
   \nobreak \vskip 3pt \nobreak}

\def\subhead#1{                 % Subhead;  NOTE enclose the text in {}
  \vskip 0.25truein             % e.g., \subhead{A. History of the Problem}
  {\raggedcenter {\it #1} \par} \nobreak \vskip 3pt \nobreak}

\def\beneathrel#1\under#2{\mathrel{\mathop{#2}\limits_{#1}}}

\def\refto#1{${\,}^{#1}$}       % For references in text as superscript

\newdimen\refskip \refskip=0pt
\def\references         % Begin references -- basic format is Ap. J., i.e.
  {\head{References}    %  [journal], [volume], [page] (space after comma).
   \beginparmode \frenchspacing \parindent=0pt \leftskip=\refskip
   \parskip=0pt \everypar{\hangindent=20pt\hangafter=1}}

\gdef\refis#1{\item{#1.\ }}                     % Ref list numbers.

\gdef\journal#1, #2, #3 {               % references,  Ap. J.  style
    {\it #1}, {\bf #2}, #3.}            % comma separates: name, vol, page

      % NB: year BEFORE journal name
%  ex:  J. Smith.  1901.  Gen. Rel. Grav.  123: 456.

% journal abbreviations:
% convention: lower case ("\prd") is Phys. Rev. style

% uppercase ("\ApJ") for Ap. J. style

\def\endreferences{\body}

\def\figurecaptions             % Begin figure captions
  {\endpage \beginparmode \head{Figure Captions}
   \parskip=3pt \everypar{\hangindent=20pt\hangafter=1} }

\def\endpage                    %  Eject a page
  {\vfill\eject}

% Ways to say goodbye
\def\endpaper   {\endmode\vfill\supereject}
\def\endjnl     {\endpaper\end}

%%
%%      Various little user definitions
%%

\def\ref#1{ref.{#1}}                    %   For inline references
\def\Ref#1{Ref.{#1}}                    %       ditto
\def\[#1]{[\cite{#1}]}
\def\cite#1{{#1}}
          %   For citation of equation numbers
        %       ditto
                       %       ditto

                     %       ditto

%\def\Fig{Figure}
%\def\Figs{Figures}
                %   PRL figure caption page
          %   Ap. J. Figure caption page
\def\(#1){(\call{#1})}
\def\call#1{{#1}}
\def\frac#1#2{{#1 \over #2}}

\def\fourth{{\frac 14}}
\def\12{{1\over2}}

\def\sla{\raise.15ex\hbox{$/$}\kern-.57em}
\def\leaderfill{\leaders\hbox to 1em{\hss.\hss}\hfill}
\def\twiddle{\lower.9ex\rlap{$\kern-.1em\scriptstyle\sim$}}
\def\bigtwiddle{\lower1.ex\rlap{$\sim$}}
\def\gtwid{\mathrel{\raise.3ex\hbox{$>$\kern-.75em\lower1ex\hbox{$\sim$}}}}
\def\ltwid{\mathrel{\raise.3ex\hbox{$<$\kern-.75em\lower1ex\hbox{$\sim$}}}}
\def\square{\kern1pt\vbox{\hrule height 1.2pt\hbox{\vrule width 1.2pt\hskip 3pt
   \vbox{\vskip 6pt}\hskip 3pt\vrule width 0.6pt}\hrule height 0.6pt}\kern1pt}
\def\tdot#1{\mathord{\mathop{#1}\limits^{\kern2pt\ldots}}}

\def\pmb#1{\setbox0=\hbox{#1}%
  \kern-.025em\copy0\kern-\wd0
  \kern  .05em\copy0\kern-\wd0
  \kern-.025em\raise.0433em\box0 }

\catcode`@=11
\newcount\r@fcount \r@fcount=0
\newcount\r@fcurr
\immediate\newwrite\reffile
\newif\ifr@ffile\r@ffilefalse
\def\w@rnwrite#1{\ifr@ffile\immediate\write\reffile{#1}\fi\message{#1}}

\def\writer@f#1>>{}
\def\referencefile{%                      Stuff to write .REF file
  \r@ffiletrue\immediate\openout\reffile=\jobname.ref%
  \def\writer@f##1>>{\ifr@ffile\immediate\write\reffile%
    {\noexpand\refis{##1} = \csname r@fnum##1\endcsname = %
     \expandafter\expandafter\expandafter\strip@t\expandafter%
     \meaning\csname r@ftext\csname r@fnum##1\endcsname\endcsname}\fi}%
  \def\strip@t##1>>{}}

\def\citeall#1{\xdef#1##1{#1{\noexpand\cite{##1}}}}
\def\cite#1{\each@rg\citer@nge{#1}}     % Variable No. of args, separated by
%%","

\def\each@rg#1#2{{\let\thecsname=#1\expandafter\first@rg#2,\end,}}
\def\first@rg#1,{\thecsname{#1}\apply@rg}       % each@ag is a general purpose
\def\apply@rg#1,{\ifx\end#1\let\next=\relax%      variable no. of arg. macro.
\else,\thecsname{#1}\let\next=\apply@rg\fi\next}% args separated by commas

\def\citer@nge#1{\citedor@nge#1-\end-}  % Check for M-N range (M and N numbers)
\def\citer@ngeat#1\end-{#1}
\def\citedor@nge#1-#2-{\ifx\end#2\r@featspace#1 % Single argument
  \else\citel@@p{#1}{#2}\citer@ngeat\fi}        % M-N range of arguments
\def\citel@@p#1#2{\ifnum#1>#2{\errmessage{Reference range #1-#2\space is bad.}
    \errhelp{If you cite a series of references by the notation M-N, then M and
    N must be integers, and N must be greater than or equal to M.}}\else%
 {\count0=#1\count1=#2\advance\count1
by1\relax\expandafter\r@fcite\the\count0,%
  \loop\advance\count0 by1\relax%         Loop from M to N
    \ifnum\count0<\count1,\expandafter\r@fcite\the\count0,%
  \repeat}\fi}

\def\r@featspace#1#2 {\r@fcite#1#2,}    % Eat spaces at beginning or end of arg
\def\r@fcite#1,{\ifuncit@d{#1}          % Cite individual reference
    \expandafter\gdef\csname r@ftext\number\r@fcount\endcsname%
    {\message{Reference #1 to be supplied.}\writer@f#1>>#1 to be supplied.\par
     }\fi%
  \csname r@fnum#1\endcsname}

\def\ifuncit@d#1{\expandafter\ifx\csname r@fnum#1\endcsname\relax%
\global\advance\r@fcount by1%
\expandafter\xdef\csname r@fnum#1\endcsname{\number\r@fcount}}

\let\r@fis=\refis                       % Save old \refis, redefine
\def\refis#1#2#3\par{\ifuncit@d{#1}%      Use two params #2 #3 to strip blank
    \w@rnwrite{Reference #1=\number\r@fcount\space is not cited up to now.}\fi%
  \expandafter\gdef\csname r@ftext\csname r@fnum#1\endcsname\endcsname%
  {\writer@f#1>>#2#3\par}}

\def\r@ferr{\endreferences\errmessage{I was expecting to see
\noexpand\endreferences before now;  I have inserted it here.}}
\let\r@ferences=\references
\def\references{\r@ferences\def\endmode{\r@ferr\par\endgroup}}

\let\endr@ferences=\endreferences
\def\endreferences{\r@fcurr=0%            Save old \endreferences, redefine
  {\loop\ifnum\r@fcurr<\r@fcount%         Loop over refnum and produce text
    \advance\r@fcurr by 1\relax\expandafter\r@fis\expandafter{\number\r@fcurr}%
    \csname r@ftext\number\r@fcurr\endcsname%
  \repeat}\gdef\r@ferr{}\endr@ferences}

% Save old \endpaper, redefine it to write parting message.

\let\r@fend=\endpaper\gdef\endpaper{\ifr@ffile
\immediate\write16{Cross References written on []\jobname.REF.}\fi\r@fend}

\catcode`@=12

\citeall\refto          % These macros will generate citations
\citeall\ref            %
\citeall\Ref            %
%%%%%%%%%%%%%%%%%%%%%%%%%%%%%%%%%%%%%%%%%%%%%%%%%%%%%%%%%%%%
%\vglue 1. truein
\title
{
Topological Defects in Cosmology\footnote{*}{Lectures delivered at ICTP,
Trieste, July 1993.}
}
\author
{Tanmay Vachaspati}
\affil
{
Tufts Institute of Cosmology,
Department of Physics and Astronomy,
Tufts University, Medford, MA 02155
}

\abstract
%\doublespace
\oneandahalfspace

The scenario of a cosmology with topological defects is surveyed
starting from the field theoretic aspects and ending with a description
of large-scale structure formation and magnetic field generation.

\endtopmatter
\oneandahalfspace

\head{TOPICS}

\

1. Introduction

2. Topological defects: field theory

3. Embedded defects: field theory

4. Cosmological formation of defects

5. Cosmological constraints on domain walls and monopoles

6. Cosmic strings: general properties

7. Cosmic strings: gravitational properties

8. Structure formation and magnetic field generation by cosmic strings

\

\beginsection{\bf {1. Introduction}}

Classical, non-dissipative solutions -  called ``lumps'' or ``defects'' - have
been studied extensively in the history of physics with various motivations
behind their investigation. Initially, they were studied in the context of
Scott Russell's waves in hydrodynamics\refto{sr},
then as ``solitons'' by mathematical
physicists and more recently by particle physicists, as possible new
``particles'' in the spectrum of non-linear field theories. They have been
investigated by condensed matter physicists looking into superconductivity and
by astrophysicists studying the formation of galaxies. Even without considering
the fluids and condensed matter research, a vast amount
of literature\refto{crgs, sc, raja, avphysrep, avps} has accummulated
on different aspects of such solutions and the interest continues unabated.

Lumps have been observed in fluids,
plasmas and a number of condensed matter systems but, in particle
physics, not a single solution of this type has been detected. The magnetic
monopole remains elusive and, in fact,
the archetypal GUT monopole is perceived to be
enough of a cosmological problem that it has to be ``inflated'' away.
Cosmic strings, though astrophysically promising, have
yet to be detected and are tightly constrained by the millisecond
pulsar and other observations. Heavy domain walls are believed to be
cosmological disasters and a particle physics model is considered inadmissible
if it predicts them. The question still looms large if there are any
classical, non-dissipative solutions in particle physics.

On the other hand,
the detection of a defect in a system can give us valuable information
about the system. Since the defect is non-perturbative,
it gives us information about the non-perturbative structure
of the theory. The existence of a topological defect would tell us
something about
the topology of the theory and with it, other features that would
be impossible to glean by perturbative scattering experiments. A famous
example is that due to Dirac\refto{dirac}:
the very existence of a monopole would
tell us that electric charge is quantized\footnote{*}{It should also be
said that this reason is no longer as compelling as it used to be in
view of the fact that electric charge is automatically quantized
within the framework of Grand Unified theories.}.
The presence of lumps in a system can lead to novel phenomenon:
GUT monopoles can catalyze proton decay\refto{rubakov, callan},
and, strings and textures
can lead to galaxy formation and to the generation of primordial
magnetic fields\refto{nt, yz, av, tvavwiggly}.
Lumps can also give rise to exotic
quantum phenomena such as fermionic zero
modes\refto{rjpr} and quantum hair on black holes\refto{scjpfw}.
The benefits that would be reaped if lumps exist in particle
physics seem to far
outweigh the doubts one may have about their existence. It is hardly
surprising, then, that so much effort has gone in the past several decades
to uncover the mysteries of the lump.

In the following notes, we shall merely touch upon certain basic aspects of
this immense subject and hope that this can be a starting point for the
reader to follow up on the references that have been provided. The choice
of topics included here are essentially aspects of the subject that I have
been personally involved in together with some basic underlying topics that
are the foundations on which the subject has grown.

\vfill
\eject

\beginsection{\bf {2. Topological defects: field theory}}

In this section we will study topological defects as classical
solutions in certain field theories\refto{kibble}.

The general criterion for the existence of a $d$ (spacetime)
dimensional topological defect
in a field theory which exhibits spontaneous symmetry
breaking from a group $G$ to a subgroup $H$ is:
$$
\pi _{3-d} (G/H) \ne 1
\eqno (2.1)
$$
where $\pi _n (G/H)$ is the $n^{\rm th}$ homotopy group of the
coset space $G/H$. The cases $d = 0, 1, 2, 3$ correspond to the
texture, monopole, string and domain wall. The condition (2.1)
leads to topological defects but the theory might also contain
non-topological, semilocal and embedded defects whose existence
cannot be detected by using (2.1).

It is often more convenient to think in terms of the vacuum manifold,
$\Sigma_V$,
described by the values of the scalar field $\phi$ that minimize the
potential:
$$
\Sigma_V = \biggl \{ \phi : {{dV} \over {d\phi }} = 0 \ ,
{{d^2 V} \over {d\phi ^2 }} > 0  \biggr \} \ .
\eqno (2.2)
$$
If the surface $\Sigma_V$ has incontractible surfaces of $3-d$ dimensions,
the theory will contain a $d$ spacetime dimensional topological defect.

A physical justification for the above criterion can be given. Suppose
that the configuration of $\phi$ on a $3-d$ dimensional surface at
spatial infinity ($S$) is denoted by
$\phi_\infty$. If we assume that there is vanishing energy at infinity,
then $\phi_\infty$ lies on $\Sigma_V$. Therefore $\phi_\infty$ describes
a mapping from $S$ to $\Sigma_V$. Next let us
imagine the case when the image of this mapping is one of the incontractible
$3-d$ dimensional surfaces in $\Sigma_V$ and denote this image by
$I$. But $S$ is contractible - assuming that space itself has
trivial topology\footnote*{If there are black holes or other gravitational
peculiarities present, these arguments would need modification.} - and so we
can continuously shrink this surface to
a point. When $S$ shrinks to a point, it must continue to be mapped
to a non-trivial surface in $\Sigma_V$ since $I$ is assumed to be
incontractible. But this would mean that $\phi$ would be multi-valued
at the point to which $S$ has been contracted and this is not acceptable.
The only way out of this contradiction is that the field $\phi$ must
leave $\Sigma_V$ at some point in space. However, this means that $\phi$
cannot remain at the minimum of the potential everywhere and there must
necessarily be at least one point where there is non-zero potential
energy. The location of this potential energy is the location of the
topological defect and the energy distribution at this location defines
the energy distribution of the defect. Note that this argument shows
that the asymptotic
configuration of the field $\phi$ is sufficient to determine the
existence of the defect.

The case of the texture is somewhat different since the asymptotic
field configuration is not sufficient to determine its presence
and at almost all times the field never leaves the vacuum manifold.
There is energy in the
configuration because the symmetries are global and hence the variations
in the field carry gradient energy. The topology can be understood by
considering the configuration in {\it spacetime} since then we can consider
{\it three spheres} in the vacuum manifold that are incontractible. If the
model one is considering only
contains gauged symmetries, there are no textures since all the gradients
in the scalar field can be compensated by gauge fields. Yet one may still
have textures in the entire universe and these result in different
sectors in the gauge theories and lead to degenerate
vacuua\refto{raja}.

To make things more concrete, we now list the simplest models that
give rise to walls, strings, monopoles and textures.

The domain wall solution arises in models in which a discrete symmetry
is spontaeously broken. For example, consider the model:
$$
S_w = \int d^4 x \left [ (\partial _\mu \phi )^2 - \lambda (\phi^2
                            - \eta ^ 2 )^2 \right ]
\eqno (2.3)
$$
where, $\phi$ is a real scalar field.
The symmetry breaking in this model is $Z_2 \rightarrow 1$ and hence
$\pi _0 (G/H) \ne 1$. In terms of the vacuum manifold, it is given
by $\phi = \pm \eta$ and consists of two disconnected minima.
The consequence is a domain wall in the model which interpolates between
the two minima.
For a static domain wall in the $yz$ plane, the field solution is:
$$
\phi = \eta {\rm tanh} ( \sqrt{\lambda}~ \eta x  )
\eqno (2.4)
$$
and the energy per unit area of a wall in the $yz-$plane is:
$$
E_w = {8 \over 3} \sqrt{\lambda} \eta^3 \ .
\eqno (2.5)
$$

The most familiar example of a model with gauge strings\refto{hnpo}
is the Abelian-Higgs model:
$$
S_s = \int d^4 x \left [ |(\partial _\mu + ie A_\mu ) \phi | ^2
               - \fourth F_{\mu \nu} F^{\mu \nu} -
{\lambda \over 4} \biggl ( \phi^2 - {{\eta ^ 2} \over 2} \biggr ) ^2 \right ]
\eqno (2.6)
$$
where, $\phi$ is a complex scalar field and
$F_{\mu \nu} = \partial_\mu A_\nu - \partial _\nu A_\mu$. Here the symmetry
breaking is $U(1) \rightarrow 1$ and $\pi_1 (G/H) \ne 1$. The vacuum
manifold is given by $\phi = {\eta \over {\sqrt{2}}} e^{i \alpha}$ where
$\alpha$ is any phase. Therefore the vacuum manifold is a circle
(parametrized by $\alpha$) and contains incontractible 1 dimensional
curves. The corresponding unit winding
string solution in cylindrical coordinates $(r, \theta , z)$, and
along the $z-$axis, is of the form:
$$
\phi = {{\eta} \over {\sqrt{2}}} f (r) e^{i  \theta }
\eqno (2.7)
$$
$$
A_\mu = - {{v (r)} \over {er}} \partial_\mu \theta
\eqno (2.8)
$$
where, the functions $f (r)$ and $v (r)$ vanish at the origin and go
to 1 as $r \rightarrow \infty$. There is no known closed form for
$f$ and $v$ and they have to be found numerically.

When $8\lambda = e^2 $, that is, when the scalar and vector masses are equal,
the energy of the vortex can be found analytically using
Bogomolnyi's method\refto{bogo} and the result is:
$$
E_s = \pi  \eta ^2 \ .
\eqno (2.9)
$$
For other values of the parameters, the energy has to be evaluated
numerically\refto{laguna}.

A simple example of the magnetic monopole\refto{thooft, ap}
occurs in the model:
$$
S_m = \int d^4 x \left [
             |(\partial _\mu + ie \epsilon^a A^a _\mu ) {\vec \phi} | ^2
                 - \fourth F^a _{\mu \nu} F^{a \mu \nu} -
{\lambda \over 4} \biggl ( {\vec \phi}^2 - {{\eta ^ 2} \over 2} \biggr ) ^2
                \right ]
\eqno (2.10)
$$
where, $\vec \phi$ is a triplet of fields, $a = 1,2,3$, $F^a_{\mu \nu}$
are the non-Abelian field strengths and
$(\epsilon ^a )_{ij} = \epsilon_{aij}$ is the usual epsilon symbols.
The symmetry breaking is $O(3) \rightarrow O(2)$ and this gives magnetic
monopoles since $\pi_2 (G/H) \ne 1$. Here the minimum of the potential
is a two sphere and hence contains incontractible two dimensional surfaces.
The monopole configuration can
be written down in spherical coordinates:
$$
\vec \phi =  {{\eta} \over {\sqrt{2}}} {{f(r)} \over {er}} {\hat r}
\eqno (2.11)
$$
$$
A^a _\mu = \epsilon_{aij} {\hat r}^i \partial_\mu {\hat r}^j
                    \biggl ( {{1-v(r)} \over {er}} \biggr )
\eqno (2.12)
$$
where,
$$
\hat r = (sin\theta ~cos\phi , sin\theta ~ sin\phi , cos\theta ).
\eqno (2.13)
$$
The functions $f(r)$ and $v(r)$ and the mass of the monopole need to be found
numerically. The only exception is in the
Prasad-Sommerfeld\refto{prasad} limit, when the
solutions and the mass of the monopole are known in closed form. This is
the limit $\lambda \rightarrow 0$ and with $e$ and $\eta$ held fixed. Then
the solution is:
$$
f(r) = Cr coth(Cr) -1
\eqno (2.14)
$$
$$
v(r) = {{Cr} \over {sinh(Cr)}}
\eqno (2.15)
$$
where, $ C = \eta e $. The energy in this limit is:
$$
E = {{4\pi \eta} \over {e}} \ .
\eqno (2.16)
$$

Finally, the simplest texture\refto{rd, ntds} occurs in a model with
symmetry breaking $O(4) \rightarrow O(3)$:
$$
S_t = \int d^4 x \left [ ( \partial _\mu \Phi ) ^2 - \lambda (\Phi^2
                            - \eta ^ 2 )^2 \right ]
\eqno (2.17)
$$
where, $\Phi$ is a column vector of 4 real fields. The vacuum manifold
in this case is a three sphere and admits incontractible three dimensional
surfaces. The texture solution is time dependent. The usual approach is
to assume spherical symmetry and only consider the $\sigma -$model limit
when $| \Phi |= \eta$. Then, in spherical coordinates\refto{ntds},
$$
\Phi = \eta (cos\chi , sin\chi \hat r )
\eqno (2.18)
$$
where, $\hat r$ is the unit radial vector as given in (2.13).
The function $\chi (t, r)$ can also be found analytically:
$$
\chi (t, r) = 2  tan^{-1} ( - u) \ , \ \ \  u < 0
\eqno (2.19a)
$$
$$
\chi (t, r) = 2  tan^{-1} ( + u)  + \pi \ , \ \ \  1 \ge u > 0
\eqno (2.19b)
$$
$$
\chi (t, r) = 2  tan^{-1} ( +1/u) + \pi \ , \ \ \  u \ge 1
\eqno (2.19c)
$$
where, $u \equiv r/t$.

This completes our synopsis of topological defect solutions. The description
is far from complete but suffices for applications to cosmology. The
reader interested in the classification of defects, defects in particle
physics models, exotic defects etc. is referred to the excellent review by
Preskill\refto{jpreview} and to the book by Vilenkin and Shellard\refto{avps}.

\vfill
\eject

\beginsection{\bf {3. Embedded defects: field theory}}

Even if the general criterion for the existence of a topological defect
(eq. (2.1)) is not satisfied, the model can still permit the existence
of topological defect like solutions\refto{tvmb, mbtvmb}. These solutions
are essentially
the topological defects of a smaller theory which are embedded in the
bigger theory under consideration. The conditions under which such
an embedding can be successfully carried out are not very stringent
and so we can expect embedded defects to exist in almost any model.

The existence of a solution in a model does not automatically mean that
it is stable and it is in this crucial way that embedded defects differ
from their topological counterparts. A single topological defect is
stabilized by topology and is separated from the zero topological
defect sector by an infinite energy barrier. Embedded defects,
however, need not be stable. In fact, mostly they are unstable,
sometimes they are metastable and the only known examples of stable embedded
defects are semilocal strings\refto{tvaa}.

Instead of giving the general arguments for constructing embedded defect
solutions, we shall only present some illustrative examples.
We shall first construct an embedded domain wall solution as this is the
simplest example of an embedded defect and then we will focus on the
electroweak model and construct the embedded electroweak string solutions.
A limiting case of electroweak strings will give us the semilocal string.

{\it Walls -} The most trivial embedded solution is a domain wall
embedded in a global $G=U(1)$ model.
We express the Higgs field in terms of two real scalar fields $\phi ^a$,
$a=1,2$. A Lagrangian that is invariant under the global $U(1)$
rotation and describes static field configurations is,
$$
L= \partial _i \phi ^a \partial ^i \phi ^a
              - \lambda \ \left( \phi^a \phi^a - \eta^2 \right )^2,
\eqno (3.1)
$$
with $i$ labeling the spatial coordinates.

The first step in constructing the
embedded domain wall solution is to identify a $Z_2$ subgroup of the full
symmetry group. Let
us consider the $Z_2$ subgroup defined by the transformation:
$(\phi_1 , \phi_2 ) \rightarrow (- \phi_1 , \phi_2 )$.
Any non-zero vacuum expectation value of
$\phi_1$ will break this $Z_2$ subgroup completely and so the
embedded symmetry breaking is $Z_2 \rightarrow 1$. This symmetry
breaking has topological domain walls:
$$
\phi_1 = \eta {\rm tanh} ( \sqrt{\lambda } ~ \eta x )
\eqno (3.2)
$$
and so this configuration for $\phi_1$
together with $\phi_2 = 0$ is our candidate embedded domain wall solution.

Once we have identified a candidate embedded defect solution, we should
check if it extremizes the energy functional. The general conditions
for this to be true can be written\refto{tvmb, mbtvmb} and require setting
up a formalism. The idea however is simple:
to check that the configuration is a solution, we
perturb the configuration and verify that the variation in the energy
vanishes to first order in the perturbations.
Since we know that the domain wall is a solution
to the theory when $\phi_2$ is zero, and the directions of $\phi_1$ and
$\phi_2$ are orthogonal, there is no need to perturb $\phi_1$ -
only perturbations in $\phi_2$ might be dangerous. Now we see that
$\phi_2$ appears quadratically in the energy functional and so
the variation in the energy functional vanishes to linear order and the
configuration is a solution.

This argument for checking when embedded configurations are solutions
can be extended to arbitrary models and defects without much difficulty.

{\it Electroweak strings\refto{yn, tvew, dvali} -}

Consider the Weinberg-Salam\refto{swas, jct} model of the electroweak
interactions. The symmetry breaking is: $SU(2)_L \times U(1)_Y \to U(1)$
and the bosonic sector of the Lagrangian is:
$$
L = - {1 \over 4} G_{\mu \nu a} G^{\mu \nu a}
    - {1 \over 4} F_{B \mu \nu} F^{B\mu \nu}
    + |D_\lambda \phi |^2 - \lambda ({\phi ^{\dag}} \phi - \eta ^2 /2 )^2 \
\eqno (3.3)
$$
where, $\phi$ is a complex doublet. The definitions of the field strengths are:
$$
G_{\mu \nu} ^a = \partial_\mu W_\nu ^a - \partial_\nu W_\mu ^a
                   + g \epsilon^{abc} W_\mu ^b W_\nu ^c
\eqno (3.4)
$$
$$
F_{\mu \nu} = \partial_\mu B_\nu - \partial_\nu B_\mu
\eqno (3.5)
$$
and the covariant derivative is:
$$
D_\mu = \partial_\mu - i {g \over 2} \tau^a W_\mu ^a -
                   i {{g'} \over 2} B_\mu
\eqno (3.6)
$$
where, $g$ and $g'$ are coupling constants and $\tau^a$ are the
Pauli spin matrices.

The electroweak energy functional follows from the Lagrangian (3.6):
$$
E = \int dz \int d^2 x \left [
          \fourth G_{ij} ^a G_{ij} ^a + \fourth F_{Bij} F_{Bij}
          + {(D_j \phi ) ^{\dag}} (D_j \phi ) +
            \lambda ({\phi ^{\dag}} \phi - \eta ^2 /2 )^2
              \right ]
\eqno (3.7)
$$
where, $i,j = 1,2,3$ and we have restricted ourselves to the case when
there is no time dependence and the time components of all gauge fields
vanish. In addition, since we will only be interested in string solutions,
we will only consider configurations that do not depend on the $z-$direction.
Then the integration over $z$ can be ignored and we can think in terms of
the energy per unit length of the string.

The first step is to choose a $U(1)$ subgroup of the full symmetry group.
We choose this to be the $U(1)$ subgroup generated by
$$
{\cal T}^3 = - cos^2 \theta_w \tau^3 + sin^2 \theta_w {\bf 1}
                   = diag(-\cos 2\theta_w , 1) \ .
\eqno (3.8)
$$
(Note that ${\cal T}^3$ is the generator corresponding to the $Z-$boson
(see (3.11) below)).

Now the candidate embedded string solution may be written down:
$$
\phi_{emb} = {\eta \over {\sqrt{2}}}
                f_{vor}(r) e^{i {\cal T}^3 \theta } \phi_0 .
\eqno (3.9)
$$
where, we take,
$$
\phi_0 = \pmatrix{0\cr 1\cr} \ .
\eqno (3.10)
$$
Here, $(r,\theta )$ are polar coordinates. In addition, we want that
the covariant derivatives vanish at infinity and so we take
$$
Z_\mu = cos \theta_w W_\mu ^3 - sin \theta_w B_\mu = [ A_\mu ]_{vor}
\eqno (3.11)
$$
where, $[A_\mu ]_{vor}$ is defined by eq. (2.8). All the other fields
in the model are taken to vanish.

One can check that the static configuration (3.9), (3.11) extremizes
the energy functional and hence is a solution\refto{tvmb}.

Different choices of the (``embedded'') subgroup lead to other string
solutions. The choice that we now consider is the subgroup sitting
entirely in the $SU(2)$ factor of the electroweak model and generated
by: ${\cal T}_\alpha \equiv sin \alpha \ \tau^1 + cos \alpha \ \tau ^2$,
where $\alpha$ is some
constant. Then the corresponding embedded string solution is:
$$
\phi_{emb} = f_{vor}(r) e^{i {\cal T}_\alpha \theta} \varphi_0 \ ,
\ \ \
sin \alpha W^1 _i + cos \alpha W^2 _i = (A_i)_{vor} \ ,
\eqno (3.12)
$$
and all other orthogonal combinations of gauge fields vanish.

The one parameter family of string solutions in (3.12) is called
the $W(\alpha )$ string since the flux in the string is purely in the
$SU(2)$ sector. Furthermore, by a global gauge transformation,
any single string solution in the family - that is, a string with any
value of $\alpha$ - may be transformed into the string configuration
with $\alpha = 0$. Explicitly, this gauge transformation is:
$$
\phi ' = exp\left [ - i {{(1+\tau ^3 )} \over 2}
                     \alpha \right ] \phi
\eqno (3.13)
$$
together with a corresponding transformation of the gauge fields.
This does not, however, mean that if there are many different string
solutions of different $\alpha$ present, they can be gauge transformed
to another multi-string configuration with all strings having the same value
of $\alpha$. The simplest way to see that $\alpha$ is a non-trivial
parameter is to consider a loop of $W$ string such that $\alpha$
runs from 0 to $2\pi$ as we go around the loop. The winding of
$\alpha$ around the loop is a discrete number and cannot be altered
by any non-singular gauge transformation. Hence, a loop with varying
$\alpha$ is not gauge equivalent to one with a constant value of
$\alpha$.

One may also see that strings of different $\alpha$ are distinct by
comparing the directions of the field strengths in group space
in each of the strings. Of course, the field strengths are only gauge
covariant and not gauge invariant and so one must first parallel transport
the gauge field of one string to the location of the other string and
then take the scalar product of the field strengths. This leads us to
the following quantity:
$$
\Delta \equiv Tr\biggl ( \tau^a F_{ij}^a ({\vec x}_2 ; \alpha )
          P \biggl [exp( - i \int^{{\vec x}_2} _{{\vec x}_1}
                 {\vec {dl}} \cdot {\vec W}^b \tau^b ) \biggr ]
                \tau^c F_{ij}^c ({\vec x}_1 ; \alpha ') \biggr )
\eqno (3.14)
$$
The quantity $\Delta$ is a gauge invariant measure of differences in
$\alpha$ between strings.

{\it Semilocal strings -}

Consider the case $g=0$ in the electroweak model. Now the symmetry
group is $SU(2)_{gl} \times U(1)_Y$ (where $gl$ stands for ``global'')
and it breaks down to a $U(1)_{gl}$ group. The vacuum manifold is
still given by the minima of the potential and is a three sphere.
However, the model continues to have the electroweak strings as
solutions. But the $Z$-string is the only gauge string since there
is only one gauge field in the model.

With the knowledge of embedded defects, the existence of the semilocal
string solution is not a surprise, but what is surprising is its stability.
The simplest case where one can explicitly check the stability is when
$\lambda = e^2$. In this particular case, the method of Bogomolnyi can
be used and it is at once obvious that the string minimizes the
energy\refto{tvaa}. A more careful
analysis\refto{mh1, mh2} reveals that the string
is only neutrally stable in this case, unstable for larger $\lambda$
and stable for smaller $\lambda$. Subsequent numerical
studies\refto{aakklptv, rl} have confirmed this result.

One way to understand the stability of the semilocal string is by
inspecting the processes by which the string solution can destabilize.
These processes necessarily require the presence of gradients of the
Higgs field for which there can be no compensating gauge fields. Hence,
unwinding requires a growth of the gradient energy but accomplishes
a decrease of the potential energy. When the coupling constant $\lambda$
is large, the potential energy is the more important piece in the
energy functional and the string prefers to unwind. If $\lambda$ is
small, however, the gradient energy required to unwind the string
is prohibitive and so the string is stable.

Topological aspects of semilocal defects have been investigated in
Ref. \cite{ggmofr, mhrhtktv} and such defects have been constructed in a
wide range of theories in Ref. \cite{jpsemilocal}. Some cosmological aspects
of semilocal strings have been investigated in Ref. \cite{kbmb}.

Next, by considering the case of small but non-zero
values of $g$, it is clear that even the $Z-$string will be metastable
for some values of parameters. A plot of the region of parameter space
where the string is stable may be found in Ref. \cite{mjlptv}.
The stability of electroweak strings (and other embedded defects) can
be considerably enhanced if there are bound states present on the
string\refto{tvrw}.

\vfill
\eject

\beginsection{\bf {4. Cosmological formation of defects}}

So far we have only discussed topological defects as classical solutions
in certain field theories. What relevance can such solutions have for
cosmology? Here we will argue that these defects would form during
phase transitions in the early universe\refto{tkformation} and, in most
cases, would
have survived until the present epoch.

Consider a cosmological phase transition occurring at some temperature
$T$. During the phase transition the Higgs field $\phi$ acquires a
vacuum expectation value so that $|\phi | = \eta$. If the vacuum
manifold is non-trivial, this still does not fix the location of
the vacuum expectation value on the vacuum manifold. For example,
if the model is the one with $Z_2$ symmetry (eqn. (2.3)), we can
have $\phi = +\eta$ or $\phi = -\eta$ at any given point in space.
Furthermore, the value that is acquired at one point is uncorrelated
with the value acquired at some other point provided the points are
separated by a distance greater than the correlation distance
$\xi$ at the phase transition. Causality necessarily implies
that $\xi < t$ where $t$ is the epoch of the phase transition.
Hence, after the phase transition, the Higgs field lies on different
points of the vacuum manifold at different spatial points. Then,
it will happen, just by chance, that the configuration of the field
on some asymptotic surface will be topologically non-trivial. (For
example, in the Abelian-Higgs model (2.6), there will be closed circuits
in space on which the phase of the Higgs field varies from 0 to 2$\pi$ as
each of the circuits is traversed.) When this happens,
a defect will necessarily be present somewhere inside the surface.

This existence proof of the formation of defects during cosmological
phase transitions relies on two facts: (1) the presence of a defect can be
determined solely by looking at the asymptotic field configuration
and, (2) the vacuum expectation value is uncorrelated on
distances larger than $t$. This mechanism for the formation of
defects is called the ``Kibble mechanism''.

Realistically, the phase transition is a thermal process and one
should study the formation of defects using statistical field theory. Less
ambitiously, one should
estimate $\xi$ by finding the thermal correlation function during
the phase transition and this has been attempted in simple models.
A rigorous estimate of, say, the density
of defects after a phase transition is not available. However, for
cosmological purposes this is not very relevant either. The point is
that even if some defects - as few as one per horizon - are produced,
the cosmological effects can be very dramatic.

Another mechanism by which defects can be produced is by
quantum mechanical nucleation of the defect during an inflationary phase
of the universe\refto{rbagav}.
A quick way to understand this phenomenon is by thinking of inflation
as a strong force that pulls apart any object. On the other hand,
the mutual attraction of monopoles to anti-monopoles, the
tension in cosmic strings and domain walls, tends to collapse these
objects. In terms of a potential, there are two regions - one when
there is no defect and the other where there is a defect-anti-defect
pair (or loop of string, or shell of domain wall) that are being
pulled apart due to inflation. These two regions are minima of an
``effective potential'' and they are separated by a potential barrier.
But then there can be quantum mechanical
tunneling from one region to another and if there are no defects to
start with, inflation can literally pull a pair out from the vacuum.
As this is a quantum mechanical effect, the number density of defects
produced in this way is small. However, in the post-inflationary universe,
our horizon is only a small patch of the universe and it may well be
that our horizon is located in the region which does contain a few defects.
If these defects are domain walls or strings, they may still be
relevant for cosmology within our universe\refto{jgav1}.

What do the defects look like on formation? The following summary is
based on numerical simulations to study the formation of topological
defects during phase transitions.

{\it Walls:} There is one infinite wall and very few smaller walls whose size
distribution is exponentially suppressed\refto{jharveyetc, tvavform}.
The fraction of wall energy that resides in the infinite wall is about 87\% .

{\it Strings:} There are a few infinite strings with close to 80\%  of the
total string length\refto{tvavform}. The remaining string is in loops with a
scale invariant distribution:
$$
dn(R) = c {{dR} \over {R^4}}
$$
where, $dn(R)$ is the number density of loops having size between
$R$ and $R+dR$ and $c \sim 6$ is a numerical coefficient determined by
simulations\refto{tvavform}.
The loops will collapse, radiate energy and disappear
but the infinite strings will survive as they are protected by topology.

{\it Monopoles:} The number density of monopoles is $\sim \xi ^{-3}$ where
$\xi$ is the correlation distance after the phase
transition\refto{jpmonopole}. For
a second order phase transition $\xi \sim T^{-1}$ and for a first
order phase transition it is given by the typical size of the bubbles
when they collide. The radius of bubbles at collision could be anywhere
between $T^{-1}$ and the Hubble distance $\sim t$.

{\it Embedded defects:}
The Kibble mechanism does not directly apply to the formation of
embedded defects since the point (1) above is not met and, in this case,
a full study of the phase transition appears necessary. However,
some guesses can be made on the basis of what is known about the formation
of topological defects. For example, if the embedded strings are metastable
they can end on monopoles. And if the probability for a string to break by
the formation of monopole and antimonopole pair is small, one would still
expect to form infinite strings. For a larger probability, however, the
length distribution of embedded string segments is expected to be
exponential\refto{tvsimulate, tvrealistic} - that is, the number of
long strings is exponentially suppressed. At this time, however, no one
has a quantitative understanding of the formation of non-topological
defects.

{\it Nucleated defects:}
For defects that nucleate in de Sitter space, it is possible to find the
wave-function for quantum fluctuations of the defect and hence the
distribution of various shapes and sizes of nucleated
defects\refto{jgav1, jgav2}.
Cosmogical consequences of such defects have also been investigated
in Ref. \cite{jgav3}.

\vfill
\eject

\beginsection{\bf {5. Cosmological constraints on domain walls and monopoles}}

Domain walls and magnetic monopoles are strongly constrained by
cosmology. Consider domain walls first\refto{dwall}.

As discussed in the previous section, an infinite domain wall will
be formed during a domain wall forming phase transition.
The infinite domain wall will move under its own tension and try to
straighten out. Immediately after the phase transition, the motion of the
wall is damped by friction but as the plasma gets diluted by
Hubble expansion, the drag decreases and eventually the motion of
the wall is effectively undamped by friction. Hubble expansion is
still important. The single domain wall in the universe cannot
disappear since it is protected by topology and would be present in
the universe if it were ever produced. (Inflation could, however, push
the domain wall outside our horizon in which case it would be
irrelevant for our observable universe.) Assuming that the domain wall
straightens out completely, its area within our horizon
is $\sim t^2$ and its mass is $\sigma t^2$ where $\sigma$ is the energy
per unit area of the wall. Therefore the energy density in the wall is
$\sim \sigma /t$ and, for the walls not to dominate the universe
today, we require that $\sigma /t < \rho$, where, $\rho$ is the matter
density at time $t$. The energy per unit area of a
domain wall is usually (for example,
in the model (2.3)) given by $\sigma = \sqrt{\lambda} \eta^3$. Assuming that
the coupling constant $\lambda$ is of order 1, and taking
$\rho \sim 3/ (32\pi Gt^2 )$ the constraint on
domain walls gives $\eta \ltwid 10 MeV$.

The scale $\eta$ is the scale at which the phase transition occurs
and hence the only permissible walls are the ones that can form
relatively late in the history of the universe. This excludes domain
walls that form at the Grand Unification scale or even at the electroweak
scale.

Next consider magnetic monopoles. There are several interesting
bounds on the number density of monopoles - each involving
different physics. Here we will only consider the cosmological bound
which in itself is quite severe\refto{jpmonopole, kolbandturner, avps}.

The cosmological bound is that magnetic monopoles should not
overclose the universe. This means that the energy density in monopoles
$\rho_m$ today should be less than the critical density of the present
universe. The smallest possible number density of monopoles at formation
($n_f$) is one per horizon at that epoch:
$$
n_f \sim {1 \over {t_f ^3}}
\eqno (5.1)
$$
After formation they quickly become non-relativistic
and from then on their energy density decays like that of matter and
redshifts as $a(t)^{-3}$ where $a(t)$ is the scale factor of the
universe. The energy density of the radiation dominated universe, however,
redshifts faster, in proportion to $a(t)^{-4}$. Therefore the
ratio of monopole energy density to critical density ($\Omega_m$)
at time $t$ is:
$$
\Omega_m (t) = {{m n_f} \over {\rho_c (t_f )}} {{a(t)} \over {a(t_f)}}
\eqno (5.2)
$$
where, $m$ is the mass of the monopole.
With $\rho_c = 3/ (32 \pi G t^2 )$, $a(t) \propto t^{1/2}$ and
$m \sim T_f$ where, $T_f$ is the temperature at the phase
transition, we find
that $\Omega_m$ becomes one at a time $t_*$ given by:
$$
t_* \sim \biggl ( {{10^{19} GeV} \over m} \biggr ) ^8 10^{-46} s.
\eqno (5.3)
$$
For GUT scale monopoles ($m \sim 10^{16} GeV$), this epoch occurs at
$10^{-22} s$ - well before the matter era. (This justifies our use of
$a(t) \propto t^{1/2}$.) Therefore, if monopoles
were formed at the GUT epoch, their energy density would completely
overwhelm our universe and would overclose it. This is not observed
and so some way has to be found to resolve this ``monopole overabundance
problem''\footnote{*}{The problem is even
more severe when one considers the various other constraints on the
present monopole flux - such as the Parker bound or the bound
coming from neutron stars\refto{kolbandturner}.}.

The monopole overabundance problem is head-on in conflict with the
philosophy of conventional GUTs which is that the electroweak and
strong forces are unified in a Grand Unified symmetry group
- a simply connected group - at an energy scale of about $10^{16} GeV$
and that there is no new force that comes into play between the
electroweak ($10^2 GeV$) and the Grand Unification scale.
Coupled with standard cosmology, this philosophy implies that
the Grand Unified symmetry group broke down to a subgroup with a
hypercharge $U(1)$ factor at the Grand Unification scale. But then
condition (2.1) is satisfied for the case of monopoles. This means
that within conventional GUTs, the monopole overabundance problem
must be confronted.

One obvious solution is to relax conventional GUTs and allow for the
possibility of a $U(1)$ factor in the Grand Unified group. Then
monopoles will never form and there will be no overabundance problem.

If one is unwilling to relax one's philosophy of Grand Unification, there is
still a cosmological solution and another particle-physics solution to
the monopole overabundance problem.
The cosmological solution is to have an inflationary
phase during or after the formation of monopoles\refto{ag}.
The inflationary phase simply dilutes the monopoles until their number
density gets so small that they cause no problem.
The particle physics solution is due to Langacker and Pi\refto{plsyp}
who consider the formation of monopoles at the GUT stage and then another
stage when the electromagnetic symmetry gets broken. At this stage the
magnetic flux of the monopoles gets confined and the monopoles get
connected by strings. The strings bring the monopoles and antimonopoles
together, they annihilate, and subsequently the electromagnetic symmetry is
restored.

\vfill
\eject

\beginsection{\bf {6. Cosmic strings: general properties}}

Here we will summarize some properties of cosmic strings.

At formation, the string network consists of closed loops and
infinite strings. It is estimated that $\sim 80 \%$ of the energy
in the string network resides in infinite strings\refto{tvavform}.

Once formed, the strings move under their own tension and try to
straighten out. This motion is damped due to the frictional force
of the ambient matter\refto{ae, mafw}
and is also slowed due to the Hubble expansion.
The frictional force is more important than the Hubble expansion
as the matter density is high. With time, however, the matter
density gets redshifted and the Hubble expansion dominates
the frictional force. The time at which the Hubble expansion
drag become comparable to that due to friction is
$t_* \sim (G\mu )^{-2} t_{Pl}$. After this time, the frictional force can
be ignored\refto{jgms}.

The motion of a string in vacuum with energy density $\mu$ is well described
by the Nambu-Goto action:
$$
S = -\mu \int d\tau d \sigma \sqrt{- g^{(2)} } \ ,
\eqno (6.1)
$$
where, $g^{(2)}$ is the determinant of the world-sheet metric defined
by
$$
g^{(2)}_{a b} = g_{\mu \nu} \partial_a x^\mu (\tau , \sigma )
                    \partial_b x^\nu (\tau , \sigma ) \ ,
\eqno (6.2)
$$
$a, b = \tau , \sigma$; $g_{\mu \nu}$ is the metric of the background
spacetime and $x^{\mu} (\tau , \sigma )$ are the coordinates of the string
world-sheet.

The Nambu-Goto action is valid
as long as the radius of curvature of the string is much larger than
the thickness of the string\refto{df, rg, kmnt}.
Also, the action does not include the
interactions of strings when they intersect. When there is a collision
of a string with another string, it leads to the phenomenon of
intercommuting\refto{ps, plcrm, kmemcr}.
In this event, the strings intersect, exchange partners and then
again move as given by (6.1).

Another factor that plays an important role in the evolution of
the string network is the gradual loss of energy from strings into
forms of radiation. An oscillating string is a time dependent solution
to a set of field equations and so one would expect that the motion
would lead to the radiation of quanta of any fields that couple with
the fields that form the string.
However, it has been shown that the radiation in these quanta is
negligible from the strings that are expected to be of cosmological
interest\refto{aetvav}. This is solely due to the fact that the curvature
and oscillation frequencies of such
strings is very small compared to particle physics scales and hence
the only radiation that can possibly be emitted is into massless
particles. A more detailed study then shows that the dominant loss
from oscillating gauge strings is to gravitational
radiation\refto{ntgradn, tvavgradn}.
A loop of size $L$ emits gravitational radiation and loses all its
energy in a time:
$$
\tau \sim {{L} \over {\Gamma G \mu}}
\eqno (6.2)
$$
where the coefficient $\Gamma$ is numerically found to be of order
100 for certain family of string loops\refto{tvavgradn}.
(In the case of global strings, the
energy loss is dominated by the emission of Goldstone
bosons\refto{rdgold, avtvgold}.)

At times later than $t_*$, the evolution of the network of strings
is governed by string tension, Hubble expansion, intercommuting and
gravitational radiation. These four factors make the evolution complicated
enough so that no one has a clear picture of what the network looks like
at any instant. Progress in this problem has relied on the results
of numerical simulations of the string evolution and, recently, an
analytical attack is also underway.

Even though the string network is not fully understood, a few features
seem to be emerging\refto{baps, dbfb, aant, proceed, tkevoln, ectkevoln}.
First of all, the network at any instant much
later than $t_*$ seems to consist of a few infinite strings (that is,
strings that traverse the whole horizon without closing up on themselves)
and a large number of tiny loops. The size distribution of the loops
is not known but the favoured guess is that the size is given by the
gravitational radiation cut-off distance: $l \sim \Gamma G \mu t$.
(Loops smaller than this evaporate in less than a Hubble time and
would probably not be significant for cosmological purposes.)
The distribution of loops in space is not known either but, since the
loops are produced at very high velocities ($v \sim 1$) one would
expect them to be distributed roughly homogenously even if they are
initially produced in a localized region of space (for which there is
some visual evidence). The long strings are not smooth but have
a lot of irregularities. The scale of the irregularities is guessed
to be the same as the size of the loops $\sim \Gamma G \mu t$ at any
time $t$. These irregularities are called ``kinks'' or ``wiggles''
in the literature and the long strings are said to be ``wiggly''.

Cosmic strings can also have the ability to carry persistent electric
currents\refto{rjpr}. Such superconducting cosmic strings\refto{ew} can
have very dramatic cosmological signatures. Other
notable varieties of strings include global, non-abelian
and Alice strings. We shall not discuss these varieties of strings but
the reader can find a description in Ref. \cite{avps}.

\vfill
\eject

\beginsection{\bf {7. Cosmic strings: gravitational properties}}

Non-superconducting topological defects interact with their environment
primarily via gravitational forces. Here we will consider the metric
of gauge strings and textures and
describe some of the known properties\refto{av300}.

We first consider the metric of a source with energy-momentum
tensor\refto{avmetric, jgott}
$$
T_\mu ^\nu = \delta (x) \delta (y) diag(\mu ,0,0,T) \ .
\eqno (7.1)
$$
With $T= \mu = \mu_0$ this is the effective energy-momentum tensor of an
unperturbed string with string tension $\mu_0$ as seen from distances much
larger than the thickness of the string. When\refto{bc, avwiggly}
$$
\mu T = \mu _0 ^2
\eqno (7.2)
$$
this also describes the energy-momentum tensor of a wiggly
string as seen by an observer who cannot resolve the wiggles on the string.

The gravitational field of the string can be found by solving the linearized
Einstein equations with $T_\mu ^\nu$ from (7.1). This gives\refto{avmetric}
$$
h_{00} = h_{33} = 4G(\mu -T) ln(r/r_0 ) ,
$$
$$
h_{11} = h_{22} = 4G(\mu +T) ln(r/r_0 ) ,
\eqno (7.3)
$$
where, $h_{\mu \nu} = g_{\mu \nu} - \eta _{\mu \nu}$ is the metric
perturbation, $r = (x^2 + y^2 )^{1/2}$ and $r_0$ is a constant of
integration.

For an unperturbed string, $T=\mu = \mu_0$ and we get
$$
h_{00} = h_{33} = 0, \ \ \
h_{11} = h_{22} = 8G\mu_0 ln(r/r_0 ) .
\eqno (7.4)
$$
A coordinate transformation brings this metric to a locally flat form,
$$
ds^2 = dt^2 - dz^2 - dr^2 - (1-8 G\mu_0 ) r^2 d\phi ^2
\eqno (7.5)
$$
It describes a conical space, which is just a Euclidean space with a wedge
of angular size $\Delta _0 = 8\pi G \mu _0$ removed and the two faces
of the wedge identified. A particle at rest with respect to a straight
string experiences no gravitational force, but if the string moves with
velocity $v_s$, then nearby matter gets a boost
$$
u_i = 4\pi G\mu _0 v_s \gamma _s
\eqno (7.6)
$$
in the direction of the surface swept out by the string. Here,
$\gamma _s = (1-v_s ^2 )^{-1/2}$. This effect is responsible for the
formation of wakes\refto{jsav}
and for a discontinuous change of the microwave
background temperature across a moving string\refto{jgott, nkas}.
Assuming that the string
is perpendicular to the line of sight, the magnitude of the latter
effect is
$$
{{\delta T} \over T} = 8\pi G\mu _0 v_s \gamma _s \ .
\eqno (7.7)
$$
The conical metric also results in the formation of double images of
background objects. In the cosmological context, this would lead to the
gravitational lensing of background quasars and
galaxies\refto{avmetric, jgott}.

Returning now to the wiggly string metric (7.3), we first consider the effect
of the wiggles on light propagation\refto{tvavwiggly}.
Assuming for simplicity that the
direction of propagation is perpendicular to the string, we can write
the relevant components of the metric in the form
$$
ds ^2 = (1+h_{00} ) [dt^2 - (dx^2 + dy^2 )]
\eqno (7.8)
$$
where we should identify the half-lines
$y=\pm 4 \pi G \mu x$, $x \ge 0$.
The conformal factor $(1+h_{00} )$ does not affect light propagation and can
be dropped.
Then the resulting metric describes Minkowski space with a deficit angle
$8 \pi G \mu$, and we conclude that background temperature
discontinuities produced by wiggly strings are given by the same equation
(7.7) with $\mu _0$ replaced by $\mu$.
In contrast to the smooth string, however, the wiggly string also produces
a change in the photon temperature when the photon propagates parallel to
the string but perpendicular to the velocity of the string.

Next, we study the formation of a wake behind a moving wiggly string.
First look at the problem in the rest frame of the string where
the particles are
flowing past the string with a velocity $v_s$ in the x-direction.
The linearized geodesic equations in the metric (8)
can be written as
$$
2 \ddot x = - ( 1- { \dot x }^2 - {\dot y} ^2 ) \partial _x h_{00} ,
\eqno (7.9)
$$
$$
2 \ddot y = - ( 1- { \dot x }^2 - {\dot y} ^2 ) \partial _y h_{00} ,
\eqno (7.10)
$$
where over-dots denote derivatives with respect to $t$.
We need only work to first order
in $G\mu$, in which case (7.10) can be integrated over the unperturbed
trajectory $x = v_s t$, $y = y_0$. Then we can transform to the frame
in which the string has a velocity $v_s$.
The result for the velocity impulse in the y-direction after the
string has passed by is\refto{tvavwiggly}:
$$
u_i = - {{2\pi G (\mu -T)} \over {v_s \gamma _s}}
             - 4 \pi G \mu v_s \gamma _s
\eqno (7.11)
$$
The second term is the usual
velocity impulse due to the conical deficit angle. But, for small
velocities, it is the first term that dominates the deflection
of particles. The origin of this term can be easily understood.
{}From eqn. (7.3), the gravitational force on a non-relativistic particle
of mass $m$ is $F = 2m G(\mu - T) /r$. A particle with an impact
parameter $r$ is exposed to this force for a time
$\Delta t \sim r/v_s$ and the resulting velocity is
$u_i \sim (F/m) \Delta t \sim G(\mu - T) / v_s$.

The metric of a loop of cosmic string can be obtained quite easily
in the weak field approximation\refto{ntloop, tvloop}.
The result is that, at distances
much larger than the size of the loop, the metric is Schwarzschild with
mass parameter $M$ equal to the mass of the loop. Singular points
(``cusps'') on the string world-sheet may have novel gravitational
features\refto{tvloop} but a proper treatment of these features requires
going beyond the weak gravitational field approximation.
(In addition, these singular features
do not seem to be very relevant for cosmology since now it is believed
that loops are not very important and that the occurrence of cusps on loops
is not as generic as it first seemed to be\refto{dgtv}.)

\vfill
\eject

\beginsection{\bf {8. Structure formation by wiggly cosmic strings}}

The formation and evolution of long-string wakes and their possible role
in structure formation in the scenario where strings are not wiggly
have been discussed in
Refs.\cite{jsav, tv, mr, asetal, thsm, lprbas} and many other papers.
The scenario where the strings
are wiggly has a shorter history and some of the relevant papers can
be found in Refs. \cite{tvavwiggly, dv1, dv2, aaas1, aaas2, tvstructure}.
Here we will follow Ref. \cite{tvavwiggly}.

When a collisionless fluid flows in the wiggly string metric, the
gravitational field focusses the fluid particles inwards so that
streamlines flowing on either side of the string converge behind the
string. Therefore a wake forms behind the string which has twice the
density of the ambient fluid. If the fluid is not collisionless, the
fluid flow into the wake will be accompanied by the formation of shocks
and turbulence. These features of the wake are likely to be important
for the formation of structure on galactic scales and for the generation
of magnetic fields. However, the details of the wake are not important
for the formation of large-scale structure which is what we shall now
describe.

Consider a wake formed behing a moving string segment of
length $\sim \xi (t_i )$ at time $t_i$. The distance travelled by the
string in one Hubble time is $\sim v_s t_i$, and thus the initial
length and width of the wake are $l_i \sim t_i$, $w_i \sim v_s t_i$.
We shall first assume that the universe is dominated by cold
dissipationless matter. In this case the two opposite streams of
matter in the wake overlap, and the mass density is enhanced by
a factor of 2 within a wedge with an opening angle $2u_i / v_s$,
where $u_i$ is from eqn. (7.11). The average thickness of this wedge
is $d_i \sim u_i t_i$. The initial surface density of the wake
is
$$
\sigma _i \approx 2 \rho (t_i ) d_i
                    \approx {2 \over 3}
                            {{\mu -T} \over {v_s t_i}}
\eqno (8.1)
$$
where $\rho (t) = (6\pi G t^2 )^{-1}$ is the average density of
the universe, and its total mass is
$M_i \approx \sigma _i l_i w_i \approx (\mu - T) t_i$. Note that
$M_i$ is independent of the string velocity $v_s$. If the string
moves faster, the wake is wider, but the surface density is
decreased proportionately. We note also that the velocity perturbation
(7.11) is produced at distances up to $\sim w_i$ from the plane of the
wake. For $r > w_i$, the gravitational field of the string is like that
of a stationary rod and $u_i \sim G(\mu - T)t_i /r$.

As the universe expands, the length and width of the wake grow like the
scale factor, $a(t) \sim t^{2/3}$, while the total mass of the wake
grows by gravitational instability like $M \propto a(t)$. As a result,
the wake thickness (defined as the turnaround distance)
and surface density evolve like $d \propto a^2 (t)$,
$\sigma \propto a^{-1} (t)$. At the present time ($t=t_0$)
$$
\sigma _0 \approx {{\mu - T} \over {v_s t_0}}
                \left ( {{t_0} \over {t_i}} \right ) ^{1/3} .
\eqno (8.2)
$$
Cold-dark-matter wakes can also be formed during the radiation era
($t_i < t_{eq}$), but in this case the gravitational instability
sets in only at $t \sim t_{eq}$. It can be shown that the surface
density of the resulting wakes is proportional to
$(t_i /t_{eq} )^{1/2}$. Together with eqn. (8.2) this implies\refto{tv}
that the most prominent wakes having the largest surface density
are the ones formed at $t \sim t_{eq}$.

The fraction of the total mass of the universe accreted onto wakes
which were formed at time $\sim t_i$ can be estimated as
(for $t_i > t_{eq}$)
$$
f \approx {{2w_i d_i z_i} \over {L^2 (t_i )}}
         \approx 8\pi G (\mu - T) z_i
\eqno (8.3)
$$
where $z_i$ is the redshift at $t_i$.
The total mass of dark matter in all wakes is dominated by the wakes
formed at $t \sim t_{eq}$,
$$
f_{tot} \sim 20 G\mu _0 z_{eq} \sim 0.4 h^2 \mu _6 .
\eqno (8.4)
$$
Here, $h$ is the Hubble constant in units of $100 km/s.Mpc$, the universe
is assumed to have critical density, $\Omega = 1$,
$\mu _6 = G\mu _0 / 10^{-6}$, and in the last step we have used the values
of $\mu$ and $T$ from the simulations.

The evolution of the initial velocity perturbation (7.11) can be found
from the equation of motion for dark matter particles,
$$
\dot u + {{\dot a} \over a} u = g
\eqno (8.5)
$$
where, $g = 2\pi G \sigma (t)$ is the gravitational acceleration due to
the wake. This gives $u(t) \propto t^{1/3}$.
A careful analysis shows that the present velocity
perturbation due to a single string impulse is\refto{thsm}
$$
u_0 \approx {2 \over 5} u_i z_i ^{1/2} .
\eqno (8.6)
$$

If the filamentary wakes created by strings were sheet-like, they
would have a thickness $\sim u_i t_i z_i ^2$.
For wakes produced around $t_{eq}$, this thickness is
somewhat larger than the width of the wake $\sim v_s t_i z_i$ and
hence we should not treat the filamentary wake as having planar
geometry. Instead the wake should be treated as having linear
geometry and the accreting structure will be cylindrical in shape
with the possibility of some planar sub-structure. The diameter of the
cylindrical structure is characterized by the geometric mean
of the previously calculated width and thickness and is
$\sim (u_i v_s )^{1/2} z_i ^{3/2} t_i$ while the length is $t_i z_i$.

To get a qualitative feel for the appearance of the wakes, we adopt the
picture developed in Ref. \cite{baps} for the evolution of the string network.
The basic idea is that the long strings are moving slowly at speeds $\sim
0.2$ for about one Hubble time. Then there is an intercommuting
somewhere in the network which triggers an instability and
speeds up the string to a much higher velocity $\sim 0.6$. In this way,
during every Hubble time period, a string moves slowly for most of the time
but the slow motion is followed by a rapid motion that helps maintain the
scaling solution in which the distance between strings stays a fixed fraction
of the horizon size. In addition, string simulations show\refto{dbfb, baps}
that the coherence length
of strings, beyond which the directions along the string are uncorrelated,
is $\xi (t) \approx t$. The inter-string separation $L(t)$ is of the
same order of magnitude. In the matter era $L(t) \approx 0.7 t$. The
rms string velocity on the scale of the smallest wiggles is\refto{dbfb}
$(<v^2 >)^{1/2} \approx 0.6$,
but the coherent velocity obtained by averaging over a scale $\xi$
is $v_s \sim 0.15$. The average mass per unit length and string
tension are (in the matter era) $\mu \approx 1.4 \mu _0$,
$T \approx 0.7 \mu _0$. With these values, the first term in
eqn. (7.11) is about ten times larger than the second.

If a string segment moves coherently
for more than one Hubble time, the resulting wake will have a variable
surface density, with denser parts being the ones formed at earlier
times. The straightening of long strings on the scale $\xi \sim t$
occurs mainly due to string intersections. Long strings occasionally
self-intersect producing a horizon-size loop which then rapidly
collapses into miriads of tiny stable loops. If two different strings
intercommute, the highly curved regions near the points of intercommuting
develop a high velocity, $v_s \sim 1$, and also shed off a large number
of tiny loops as they move. The wakes due to rapidly moving strings
have the form of sheets with dimensions
$t_i z_i \times t_i z_i \times u_i t_i z_i ^2$
while the wakes due to slow strings have a filamentary appearence.
As we explained (see below (8.1)), the masses of both types of wakes are
comparable, but the surface density in the filamentary wakes is much higher,
and we expect filamentary features to be prominent in the large-
scale galaxy distribution. In addition to wakes due to long strings,
there will also be comet-like wakes produced by rapidly moving small
loops\refto{eb1}. The characteristic
scale of the large-scale structure in this scenario is
$t_{eq} z_{eq} \sim 10 h^{-2} Mpc$. With $h=0.5$ it is comparable
to the scale suggested by observations\refto{cfa} ($\sim 25 h^{-1} Mpc$).

The wiggliness of the string network implies that the wakes will not
be uniform but will have sub-structure on the scale of the wiggles.
This scale is expected to be larger than the damping scale due
to gravitational radiation from the string network which
is $\Gamma G \mu t$, at the time of formation of the wake,
where $\Gamma \sim 10^2$
is a numerical factor coming from the rate of gravitational
radiation\refto{tvavgradn}. For the wakes produced at $t_{eq}$, the
sub-structure is on a comoving scale larger than
$\sim 1 \mu_6 h^{-2} kpc$.
We expect that the wakes will fragment into smaller objects
due to this sub-structure.

The large-scale velocities predicted at the present time can be found
from eqn. (8.6). For sheet-like wakes from rapidly moving strings,
it gives $u_0 \sim 300 \mu _6 \ h \ km/s$
where we assumed that $t_i \sim t_{eq}$ and
$v_s \gamma _s \approx 1$. These velocity perturbations
extend over regions of size $( 10 h^{-2} Mpc )^3$ and may account
for the observed large-scale streaming velocities\refto{dbetal}.
Reasonable values of $u_0$ are obtained, e.g., for $h \sim 0.5$,
$\mu _6 \sim 4$.
We note that in some regions of space the motion of matter can be
affected by two or more different strings. The streaming velocity
in such regions will typically be enhanced by a factor $\sqrt{n}$
where $n$ is the number of string impulses that the matter
experiences\refto{tvpeculiar}.
Regions larger than $( 10 h^{-2} Mpc )^3$ will also get peculiar
velocities due to string impulses but the velocity will scale as
$1/L$ where $L$ is the size of the region. (This is simply because
a string gives a coherent impulse to a region of size $L$, when
its correlation length becomes comparable to $L$. For large $L$,
this happens later, giving less time for the velocity to grow.)
The observational situation on the dependence of peculiar velocity
on length scale is quite unclear and it remains to be seen if this
predicted fall-off agrees with observation\refto{lptvpeculiar}.

If the dark matter is cold, wakes formed at all epochs prior to
radiation-matter equality will survive and density fluctuations will
be present on very small scales too. Albrecht and Stebbins\refto{aaas1,
aaas2} argue that this small scale power is excessive and the sheet-like
structures formed later would not be prominent. The situation to me
does not seem as clear since one could imagine small scale structures
themselves clustering into larger scale structures. So the large scale
wake could be prominent simply because it rearranges the small scale
structure into sheets and filaments.

In a universe dominated by light neutrinos, wake perturbations are
damped by neutrino free streaming on co-moving scales smaller than
$\lambda _\nu (t) \sim v_\nu (t) t$, where
$v_\nu (t) \approx v_{eq} (t_{eq} /t) ^{2/3}$ is the rms velocity
of neutrinos and $v_{eq} \approx 0.2$. On larger scales the evolution
of perturbations is similar to that in cold dark matter. For a
cold-dark-matter wake formed at time $t_i$, all matter initially
within a distance $u_i t_i z_i/z$ will be accreted onto the wake
by the redshift $z$. A neutrino wake will go nonlinear
at the redshift $z_{nl}$ when
the co-moving scale of $\lambda_\nu (t_i )$ becomes less than the
distance to which the matter has been swept by the wake\refto{asetal} :
$u_i t_i ( a_{nl} / a_i ) ^2 \approx \lambda _\nu (t_i ) a_{nl} /a_i$,
where the scale factor $a(t)$ is related to the redshift by
$1+z = a(t_0 )/a(t)$. For filamentary wakes,
this gives\refto{com2}
$1 + z_{nl} \approx 4.5 \mu_6 h^2$ independent of $z_i$. With
$h = 0.5$ and $\mu _6 = 4$, we have $z_{nl} \approx 3.5$.
For sheet-like wakes, we find $1+ z_{nl} \approx 2 \mu_6 h^2$.
Observations do indicate that $z = 2-3$ is the epoch of intensive
galaxy and quasar formation\refto{jpjs}.
Thin wakes of small relativistic loops are strongly suppressed
by the neutrino free streaming\refto{eb2}, and it appears that
loops play a negligible role in this scenario. Eqn. (8.4) then implies
that most of the matter in the universe remains unclustered at the
present time\refto{lprbas}. This may explain why dynamical
measurements in clusters give values of $\Omega$ substantially
smaller than 1. By contrast, in the cold dark matter scenario
the loops accrete at least as much matter as the wakes, and the
voids will be pierced by the long comet-like wakes formed behind
relativistically moving loops.
The characteristic
scale of the large-scale structure in this scenario is
$t_{eq} z_{eq} \sim 10 h^{-2} Mpc$. With $h=0.5$ it is comparable
to the scale suggested by observations\refto{cfa} ($\sim 25 h^{-1} Mpc$).
The surface density of the neutrino wakes produced subsequently decreases
but the decrease is only $\propto t_i ^{-1/3}$. This means that the
structure on still larger scales can also be prominent.

Baryonic wakes in a neutrino-dominated universe start collapsing
after baryons decouple from radiation, $t > t_{dec}$. However,
since baryons constitute only a small fraction of the total density,
the growth of these wakes is strongly suppressed. Baryonic wakes
could nonetheless be cosmologically significant if the energy
output from the primordial stars formed in the wakes might trigger
some kind of explosive amplification and lead to preferential
galaxy formation along these wakes\refto{mr}. They could also
explain the existence of quasars at redshifts greater than 3.
The scale of baryonic
wakes, $t_{dec} z_{dec} \sim 50 h^{-1} Mpc$, is comparable to the
largest-scale structure observed in the universe.

A novel outcome of the cosmic string scenario is that it predicts
the generation of primordial magnetic
fields\refto{tvavwiggly, tvstructure, dv3}. The mechanism by
which this happens relies on the fact that the relativistic motion
of strings after decoupling of matter and radiation induces
vorticity in the baryonic fluid. The vorticity then leads to
the generation of primordial magnetic fields.

To put these arguments on a firmer footing it is necessary to
establishe generation of vorticity and then to show that the
vorticity will lead to magnetic fields. Fortunately, the second
step had been investigated in the 70's and several mechanisms
are known by which vorticity can lead to magnetic fields. So the
main task that remains is to show that there will be vortical
motion and to estimate the vorticity.

An estimate of the Reynold's number for the flow of the baryonic fluid
into the string wake shows that it is very high ($\sim 10^{11}$) and
hence it is natural to suspect that the flow will be turbulent.
A large Reynold's number, however, is not sufficient to guarantee
turbulence and one must demonstrate that the flow is unstable. In the
case of fluid flow into a cosmic string wake one can actually go further
and explicitly describe the mechanism by which vorticity is generated.

Consider the wake formed by a relativistically moving string at the
recombination epoch. At this epoch, the sound speed is dropping steeply
but the flow of the fluid into the wake is still given by (7.11). Then,
for string tensions that are suitable for structure formation, the
fluid flow is supersonic and the wake is bounded by strong shocks. On
large-scales this shock is uniform but on small scales, the shock is
non-uniform because the string is wiggly and the wiggles have highly
variable velocities. This is the crucial feature - the wiggly string
wakes are bounded by {\it strong, non-uniform shocks}.

Once we have shown that the scenario has this feature, the presence
of vorticity follows. From Euler's equation and steady flow, one
finds that the vorticity is related to the gradient of the entropy
by the equation\refto{landau}:
$$
\vec v \times \vec \omega  = - T \vec \nabla s
\eqno (8.7)
$$
where, $\vec v$ is the fluid velocity, $\omega$ the vorticity, $T$
the temperature and $s$ the specific entropy.
In the preshock region, the flow is isentropic but, at the shock, the
entropy suffers a discontinuous jump. Since the shock is non-uniform,
the post-shock entropy is different at different points along the shock.
This gives us gradients in the post-shock entropy and a non-zero vorticity.

The presence of vorticity in the baryonic fluid flow means that the
protons and electrons are in vortical motion. But we also have ambient
photons and neutral particles which interact with the protons and electrons.
The next crucial ingredient in the scenario is that the masses of the
protons and electrons are different and therefore their interaction
times with photons are also different. (The scattering cross-section
of either particle with photons is inversely proportional to the square of
the mass of the particle.) Then the stronger interaction of photons with
electrons slows them down with respect to the protons and the resulting
electric current due to the differential rotation of charges produces
a magnetic field. Such scenarios - using the different interaction rates
of protons and electrons with photons - were proposed by Harrison\refto{eh},
Mishustin and Ruzmaikin\refto{mishustin} and others in the 70's.

While the scenario is clear qualitatively, quantitative estimates are
more difficult to obtain. The first hurdle is to understand the vorticity
in the fluid flow. For this we would need to have a quantitative
analysis of the turbulence in the fluid flow and, as far as I know, there
is no theoretical recipe for analyzing turbulent flow. But dimensional
arguments allow us to estimate the average vorticity as follows.
We expect that vorticity will be produced on the scale on which the
wake is inhomogenous. Therefore the
co-moving scale of this vorticity is $\sim \Gamma G \mu t_i z_i$
where $z_i$ is the redshift at the epoch $t_i$ when the vorticity
is generated at a time $t_i$. The velocity of the fluid is estimated
as in (7.11) and so the vorticity is
$$
\omega \sim {u \over l} \sim {{0.1} \over {t_i}} \ .
\eqno (8.8)
$$
Given the vorticity, the generated magnetic field is estimated from the
results of Mishustin and Ruzmaikin\refto{mishustin}
$$
B \approx 2 {{mc^2} \over e} {{(1+z)^{5/2}} \over
                    {\Omega ^{1/2} H \tau_{e\gamma (0)}} } \omega
\eqno (8.8)
$$
where, $m$ is the electron mass, $e$ the electron charge, $z$ the red-
shift at which the vortical motion starts, $\Omega$ the mean-to-critical
density or the baryonic matter, $H$ the Hubble constant,
$\tau _{e\gamma} (0)$ the interaction time between electrons and photons
at the present epoch and $\omega$ is the angular velocity of the
eddy. The interaction time between electrons and photons at the
present epoch is given by
$$
\tau _{e\gamma} ^{-1} (0) = {{4\sigma _T \rho _\gamma (0) c } \over {3m}}
\eqno (8.9)
$$
where, $\sigma _T \sim 10^{-24} \ cm^2$ is the Thompson scattering
cross-section and $\rho _\gamma (0)$ is the present photon energy
density.
Inserting $\Omega \sim 0.03$, $H \sim 50 \  km/s-Mpc$, $z = 10^3$,
and $\omega = 10^{-13} \ s $ gives,
$$
B \sim 10^{-14} \ G \ .
\eqno (8.10)
$$
The magnetic field produced due to the vorticity at decoupling
can be further amplified by turbulence in the wake and by a galactic
dynamo. Such a tiny seed field is all that is needed to generate the
observed galactic magnetic field of $10^{-6} G$.

Cosmic strings seem to be uniquely suited for generating magnetic fields
via vorticity since they naturally have the two features that seem
essential for this mechanism: coherence and strength.
The first feature is that the vorticity should be
on relatively large scales ($\sim 10 kpc$) for which it is essential that
the source producing the vorticity should also have this coherence scale.
If the source is not lineal , it is
difficult to see how this large a scale is obtained. The second point is
that the vorticity on these scales has to be relatively large, implying
that the source itself has to be undergoing violent motion. Once again,
this feature comes up naturally in the string picture but seems hard to
get with other sources.

{\it Outlook:}

While the cosmic string scenario for structure formation has
ingredients that seem to be promising, it is not detailed enough yet
to be testable. This is because the problem is doubly difficult - first
one has to understand the evolution of the string network and then the
evolution of the matter that gravitates around the network. The evolution
of the network itself has turned out to be a really sticky problem
and an analytical understanding is just beginning to emerge. The flow
of matter around the string network promises to be an even more difficult
problem since, as we have seen while discussing the generation
of magnetic fields, the flow will be non-linear and turbulent. Without
going into
the details of the flow, one can only make some broad predictions about
the {\it large-scale} structure as we have done above. To make predictions
on galaxy scales, it is essential to understand the structure and
fragmentation of the wake. Only then will we be able to say something about
the galaxy-galaxy correlation function and other quantitative measures that
can enable a comparison with observation.

These problems seem so difficult that at times I am tempted to think that
it may be simpler to observe cosmic strings or rule them out (as a mechanism
for structure formation) on the grounds that they are not observed. The
millisecond pulsar observations seem to be a foolproof way to go but
for this we have to wait for another decade or so. The microwave background
anisotropy measurement by COBE does not confirm or rule out strings. But
small-scale measurements - when they become more reliable - could test the
cosmic string scenario\footnote{*}{The situation is muddied by the
interfering possiblities of having anisotropies due to inflation (and other
sources) in addition to those due to strings.}.
The direct observation of strings by
their gravitational lensing property is also possible but is likely to be
effort consuming. On the other hand, this effort seems
very worthwhile considering all the exciting outcomes!

\

\vskip 0.5 truein

\noindent {$Acknowledgements:$}

I am grateful to Qaisar Shafi and the ICTP for the opportunity to
visit and lecture. This work was supported in part by the National Science
Foundation.

\vfill
\eject

\references

\refis{landau} L. D. Landau and E. M. Lifshitz, ``Fluid Mechanics'',
Pergamon Press (1987).

\refis{tvpeculiar} T. Vachaspati, Phys. Lett B{\bf 282}, 305 (1992).

\refis{lptvpeculiar} L. Perivolaropoulos and T. Vachaspati, Ap. J. Lett.,
to be published (1994).

\refis{dvali} Topological strings in extensions of the standard model
have been constructed in
G. Dvali and G. Senjanovi\'c, Phys. Rev. Lett. {\bf 71}, 2376 (1993).

\refis{dgtv} D. Garfinkle and T. Vachaspati, Phys. Rev. D {\bf 36},
2229 (1987).

\refis{lprbas} L. Perivolaropoulos, R. Brandenberger and A. Stebbins,
Phys. Rev. D{\bf{41}}, 1764 (1990); Int. J. Mod. Phys. A{\bf{5}},
1633 (1990).

\refis{eb2} E. Bertschinger in Ref. \cite{proceed}.

\refis{jpjs} P. J. E. Peebles and J. Silk, Nature {\bf{346}}, 233 (1990).

\refis{com2} Our estimate of $z_{nl}$ was derived assuming planar
geometry and applies only if $w > d$ at $z \sim z_{nl}$. This gives
the condition $z_i < z_* \equiv (v_s /v_{eq} ) z_{eq}$.
For $z_i > z_*$, the expression for $(1+z_{nl} )$ acquires an
additional factor $(z_* /z_i )$. We ignore this modification, since
it affects only the filamentary wakes in the narrow interval
$z_{eq} > z_i > 0.75 z_{eq}$.

\refis{asetal} A. Stebbins, S. Veeraraghavan, R. Brandenberger, J. Silk
and N. Turok, Ap. J. {\bf{322}}, 1 (1987).

\refis{dbetal} D. Burstein et. al. in ``Galaxy Distances and
Deviations from the Hubble Flow'', eds. B. Madore and R. Tully
(Dordrecht: Reidel, 1983); C. Collins, R. Joseph and N. Robertson,
Nature {\bf{320}}, 506 (1986).

\refis{cfa} V. de Lapparent, M. Geller and J. Huchra,
Ap. J. Lett. {\bf{302}}, L1 (1986).

\refis{eb1} E. Bertschinger, Ap. J. {\bf{316}}, 489 (1987).

\refis{tvstructure} T. Vachaspati, Phys. Rev. D{\bf 45}, 3487 (1992).

\refis{dv1} D. N. Vollick, Phys. Rev. D {\bf 45}, 1884 (1992).

\refis{dv2} D. N. Vollick, Ap. J. {\bf 397}, 14 (1992).

\refis{dv3} D. N. Vollick, Phys. Rev. D {\bf 48}, 3585 (1993).

\refis{aaas1} A. Albrecht and A. Stebbins, Phys. Rev. Lett. {\bf 68},
2121 (1992).

\refis{aaas2} A. Albrecht and A. Stebbins, Phys. Rev. Lett. {\bf 69},
2615 (1992).

\refis{ntloop} N. Turok, Phys. Lett. B{\bf 123}, 387 (1983).

\refis{tvloop} T. Vachaspati, Phys. Rev. D{\bf 35}, 1767 (1987).

\refis{tvavwiggly} T. Vachaspati and A. Vilenkin, Phys. Rev. Lett. {\bf 67},
1057 (1991).

\refis{jsav} J. Silk and A. Vilenkin, Phys. Rev. Lett. {\bf{53}}, 1700
(1984).

\refis{nkas} N. Kaiser and A. Stebbins, Nature {\bf{310}}, 391 (1984).

\refis{bc} B. Carter, Phys. Rev. D{\bf{41}}, 3886 (1990).

\refis{avwiggly} A. Vilenkin, Phys. Rev. D{\bf{41}}, 3038 (1990).

\refis{jgott} J. R. Gott, Ap. J. {\bf{288}}, 422 (1985).

\refis{ew} E. Witten, Nucl. Phy. B{\bf 249}, 557 (1985).

\refis{tkevoln} T. W. B. Kibble, Nucl. Phys. B{\bf 252}, 227 (1985).

\refis{ectkevoln} D. Austin, E. Copeland and T. W. B. Kibble, Phys. Rev.
D{\bf 48}, 5594 (1993).

\refis{baps} B. Allen and E. P. S. Shellard, Phys. Rev. Lett. {\bf 64},
119 (1990). See also the papers by Allen and Shellard in Ref. \cite{proceed}.

\refis{dbfb} D. P. Bennet and F. R. Bouchet, Phys. Rev. Lett. {\bf{60}}
257 (1988). See also the papers by Bennett and Bouchet in
Ref. \cite{proceed}.

\refis{aant} A. Albrecht and N. Turok, Phys. Rev. D{\bf 40}, 973 (1989).

\refis{proceed} G. W. Gibbons, S. W. Hawking and T. Vachaspati, ``The
Formation and Evolution of Cosmic Strings'', Cambridge University Press,
(1990).

\refis{rdgold} R. L. Davis, Phys. Rev. D{\bf 32}, 3172 (1985).

\refis{avtvgold} A. Vilenkin and T. Vachaspati, Phys. Rev. D{\bf 35}, 1138
(1987).

\refis{tvavgradn} T. Vachaspati and A. Vilenkin, Phys. Rev. D{\bf 31},
3052 (1985).

\refis{ntgradn} N. Turok, Nucl. Phys. B{\bf 242}, 520 (1984).

\refis{aetvav} A. E. Everett, T. Vachaspati and A. Vilenkin, Phys. Rev. D
{\bf 31}, 1925 (1985).

\refis{ps} E. P. S. Shellard, Nucl. Phys. B{\bf 282}, 624 (1987).

\refis{plcrm} R. Matzner, Computers in Physics {\bf 2}, 51 (1988).

\refis{kmemcr} K. J. M. Moriarty, E. Myers, and C. Rebbi, in ``Cosmic
Strings: The Current Status'', eds. F. Accetta and L. M. Krauss,
World Scientific, Singapore (1988).

\refis{df} D. Forster, Nucl. Phys. B{\bf 81}, 84 (1974).

\refis{rg} R. Gregory, Phys. Lett. B{\bf 206}, 199 (1988).

\refis{kmnt} K. Maeda and N. Turok, Phys. Lett. B{\bf 202}, 376 (1988).

\refis{jgms} J. Garriga and M. Sakelleriadou, Phys. Rev. D{\bf 48},
2502 (1993).

\refis{ae} A. E. Everett, Phys. Rev. D{\bf 24}, 858 (1981).

\refis{mafw} M. G. Alford and F. Wilczek, Phys. Rev. Lett. {\bf 62},
1071 (1989).

\refis{plsyp} P. Langacker and S-Y. Pi, Phys. Rev. Lett. {\bf 45}, 1 (1980).

\refis{ag} A. H. Guth, Phys. Rev. D{\bf 23}, 347 (1981).

\refis{jpmonopole} J. Preskill, Phys. Rev. Lett. {\bf 43}, 1365 (1979);
also, see the contribution by J. Preskill in Ref. \cite{ggshss};
Y. B. Zeldovich and M. Y. Khlopov, Phys. Lett. B{\bf 79}, 239 (1978).

\refis{kolbandturner} E. W. Kolb and M. S. Turner, ``The Early Universe'',
Addison Wesley (1990).

\refis{dwall} Y. B. Zeldovich, I. Yu. Kobzarev and L. B. Okun,
Zh. Eksp. Teor. Fiz. {\bf 67}, 3 (1974) [Sov. Phys. JETP {\bf 40},
1 (1975)].

\refis{avphysrep} A. Vilenkin, Phys. Rep. 121, 263 (1985).

\refis{jgav3} J. Garriga and A. Vilenkin, Phys. Rev. D{\bf 47}, 3265 (1993).

\refis{tvsimulate} T. Vachaspati, Phys. Rev. D{\bf 44}, 3723 (1991).

\refis{tvrealistic} R. Holman, S. Hsu, T. Vachaspati and R. Watkins, Phys.
Rev. D{\bf 46}, 5352 (1992).

\refis{jharveyetc} J. A. Harvey, E. W. Kolb, D. B. Reiss and S. Wolfram,
Nucl. Phys. B{\bf 201}, 16 (1982).

\refis{tvavform} T. Vachaspati and A. Vilenkin, Phys. Rev. D{\bf 30},
2036 (1984).

\refis{jgav1} J. Garriga and A. Vilenkin, Phys. Rev. D{\bf 44}, 1007
(1991).

\refis{jgav2} J. Garriga and A. Vilenkin, Phys. Rev. D{\bf 45}, 3469
(1992).

\refis{rbagav} R. Basu, A. H. Guth and A. Vilenkin, Phys. Rev. D{\bf 44},
340 (1991).

\refis{tkformation} T. W. B. Kibble, J. Phys. A{\bf 9}, 1387 (1976);
Phys. Rep. {\bf 67}, 183 (1980).

\refis{tvrw} T. Vachaspati and R. Watkins, Phys. Lett. B{\bf 318}, 163
(1993).

\refis{mhrhtktv} M. Hindmarsh, R. Holman, T. Kephart and T. Vachaspati,
Nucl. Phys. B{\bf 404}, 794 (1993).

\refis{aakklptv} A. Ach\'ucarro, K. Kuijken, L. Perivolaropoulos and
T. Vachaspati, Nucl. Phys. B{\bf 388}, 45 (1992).

\refis{rl} R. Leese and T. Samols, Nucl. Phys. B{\bf 396}, 639 (1993).

\refis{swas} S. Weinberg, Phys. Rev. Lett. {\bf 19}, 1264 (1967);
A. Salam in ``Elementary Particle Theory'', ed. N. Svarthholm, Stockholm:
Almqvist, Forlag AB, pg 367.

\refis{jct} J. C. Taylor, ``Gauge Theories of Weak Interactions'',
Cambridge University Press, 1976.

\refis{bogo} E. B. Bogomol'nyi, Sov. J. Nucl. Phys. {\bf{24}}, 449 (1976).

\refis{hnpo} H. B. Nielsen and P. Olesen, Nucl. Phys. B{\bf{61}}, 45 (1973).

\refis{scjpfw} S. Coleman, J. Preskill and F. Wilczek,
Phys. Rev. Lett. {\bf 67}, 1975 (1991).

\refis{yn} Y. Nambu, Nucl. Phys. B{\bf 130}, 505 (1977).

\refis{kbmb} K. Benson and M. Bucher, Nucl. Phys. B{\bf 406}, 355 (1993).

\refis{mh1} M. Hindmarsh, Phys. Rev. Lett. {\bf 68}, 1263 (1992).

\refis{mh2} M. Hindmarsh, Nucl. Phys. {\bf B392}, 461 (1993).

\refis{jpsemilocal} J. Preskill, Phys. Rev. {\bf D46}, 4218 (1992).

\refis{ggmofr} G. W. Gibbons, M. Ortiz, F. Ruiz-Ruiz and T. Samols, Nucl. Phys.
B{\bf 385}, 127 (1992).

\refis{tvaa} T. Vachaspati and A. Ach\'ucarro, Phys. Rev. D{\bf 44},
3067 (1991).

\refis{tvmb} T. Vachaspati and M. Barriola, Phys. Rev. Lett. {\bf 69},
1867 (1992).

\refis{mbtvmb} M. Barriola, T. Vachaspati and M. Bucher, Phys. Rev. D,
to be published (1994).

\refis{jpreview} J. Preskill in ``Architecture of the Fundamental
Interactions at Short Distances'', eds. P. Ramond and R. Stora, North-
Holland, Amsterdam (1987).

\refis{rd} R. L. Davis, Phys. Rev. D{\bf 35}, 3705 (1987); {\bf 36}, 997
(1987).

\refis{ntds} N. Turok and D. N. Spergel, Phys. Rev. Lett. {\bf 64}, 2736
(1990).

\refis{prasad} M. K. Prasad and C. M. Sommerfeld, Phys. Rev. Lett.
{\bf 35}, 760 (1975).

\refis{thooft} G. 't Hooft, Nucl. Phys. B{\bf 79}, 276 (1974).

\refis{ap} A. M. Polyakov, JETP Lett. {\bf 20}, 194 (1974).

\refis{laguna} P. Laguna-Castillo and R.A. Matzner, Phys. Rev. D{\bf 36},
3663 (1987).

\refis{kibble} T. W. B. Kibble in Ref. \cite{tkformation}.

\refis{rjpr} R. Jackiw and P. Rossi, Nucl. Phys. B{\bf 190} [FS3],
681 (1981).

\refis{nt} N. Turok, Phys. Rev. Lett. {\bf {63}}, 2625 (1989).

\refis{rubakov} V. A. Rubakov, Nucl. Phys. B{\bf 203}, 311 (1982).

\refis{callan} C. G. Callan Jr., Phys. Rev. D{\bf 26}, 2058 (1982).

\refis{dirac} P. A. M. Dirac, Phys. Rev. {\bf 74}, 817 (1948).

\refis{sr} J. Scott Russell, ``Report on waves'', Proc. of the British
Association for the Advancement of Science, London, 311 (1845).

\refis{crgs} C. Rebbi and G. Soliani, ``Solitons and Particles'',
World Scientific (1984).

\refis{sc} S. Coleman, ``Aspects of Symmetry'', Cambridge University
Press (1985).

\refis{raja} R. Rajaraman, ``Solitons and Instantons'', North-Holland (1987).

\refis{avps} A. Vilenkin and E. P. S. Shellard, ``Cosmic Strings and
Other Topological Defects'', Cambridge University Press (1994).

\refis{yz} Ya. B. Zeldovich, M.N.R.A.S. {\bf{192}}, 663 (1980).

\refis{av} A. Vilenkin, Phys. Rev. Lett. {\bf{46}}, 1169 (1981);
{\bf{46}}, 1496 (E) (1981).

\refis{tv} T. Vachaspati, Phys. Rev. Lett. {\bf{57}}, 1655 (1986).

\refis{mr} M. J. Rees, M. N. R. A. S. {\bf{222}}, 21p (1986).

\refis{mishustin} I. N. Mishustin and A. A. Ruzmaikin, Zh. Eksp. Teor.
Fiz. {\bf 61}, 441 (1971) [Sov. Phys. JETP {\bf 34}, 233 (1971)].

\refis{thsm} T. Hara and S. Miyoshi, Prog. Theor. Phys. (Japan) {\bf{81}},
1187 (1990).

\refis{avmetric} A. Vilenkin, Phys. Rev. D{\bf{23}}, 852 (1981).

\refis{eh} E. R. Harrison, M.N.R.A.S. {\bf{147}}, 279 (1970).

\refis{tvew} T. Vachaspati, Phys. Rev. Lett. {\bf 68}, 1977 (1992);
{\bf 69}, 216(E) 1992); Nucl. Phys. B{\bf 397}, 648 (1993).

\refis{ggshss} G. W. Gibbons, S. W. Hawking and S. T. C. Siklos,
``The Very Early Universe'', Cambridge University Press (1983).

\refis{av300} For a review, see A. Vilenkin in Ref. \cite{300}.

\refis{300} S. W. Hawking and W. Israel, ``300 Years of Gravitation'',
Cambridge University Press (1987).

\refis{mjlptv} M. James, L. Perivolaropoulos and T. Vachaspati, Nucl.
Phys. B{\bf 395}, 534 (1993).

\endreferences

\vfill
\eject

\endjnl
\end

\head {Figure Captions}

\item {1.} An example of an intercommuting event.

\item {2.} A picture of the string network taken from the simulation by
Allen and Shellard.

\item {3.} Collisionless fluid flow in a wiggly string background.

%\vfill
%\eject

\endjnl

\end